\documentclass[aps,twocolumn,nofootinbib,preprintnumbers,groupedaddress,letterpaper]{revtex4}

\usepackage[dvips]{color}
\usepackage[normalem]{ulem}
\usepackage{amsmath}
\usepackage{enumerate}
\usepackage{amsfonts}
\usepackage{epsfig}
\usepackage{yfonts}

\newcommand\comment[1]{}

\newcommand\poincare{Poincar\' e }

\newcommand\ov{\over }

\def\le{\left}
\def\ri{\right}

\def\({\left(}
\def\){\right)}
\def\[{\left[}
\def\]{\right]}
\def\<{\langle}
\def\>{\rangle}

\newcommand\half{{\ensuremath{\frac{1}{2}}}}
\newcommand\p{\ensuremath{\partial}}

\newcommand\field[1]{{\ensuremath{\mathbb{{#1}}}}}

\newcommand{\RR}{\field{R}}

\newcommand{\ZZ}{\field{Z}}

\newcommand{\be}{\begin{equation}}
\newcommand{\ee}{\end{equation}}
\newcommand{\bea}{\begin{eqnarray}}
\newcommand{\eea}{\end{eqnarray}}
\newcommand{\bwt}{\begin{widetext}}
\newcommand{\ewt}{\end{widetext}}

\newcommand{\bi}{\begin{itemize}}
\newcommand{\ei}{\end{itemize}}
\newcommand{\ben}{\begin{enumerate}}
\newcommand{\een}{\end{enumerate}}
\newcommand{\bca}{\begin{cases}}
\newcommand{\eca}{\end{cases}}
\newcommand{\bln}{\begin{align}}
\newcommand{\eln}{\end{align}}
\newcommand{\bst}{\begin{split}}
\newcommand{\est}{\end{split}}

\begin{document}

\title{Segmented strings from a different angle}

\author{David Vegh}
\email{dvegh@cmsa.fas.harvard.edu}

\affiliation{\it  CMSA, Harvard University, Cambridge, MA 02138, USA   }

\date{\today}

\begin{abstract}

Segmented strings in flat space are piecewise linear classical string solutions. Kinks between the segments move with the speed of light and their worldlines form a lattice on the worldsheet.  This idea can be generalized to AdS$_3$ where the embedding is built from AdS$_2$ patches. The construction provides an exact discretization of the non-linear  string equations of motion.

This paper computes the area of segmented strings using cross-ratios constructed from the kink vectors. The cross-ratios can be expressed in terms of either left-handed or right-handed variables.
The string equation of motion in AdS$_3$ is reduced to that of an integrable time-discretized relativistic Toda-type lattice.
Positive solutions yield string embeddings that are unique up to global transformations.
In the appendix, the \poincare target space energy is computed by integrating the worldsheet current along  kink worldlines and a formula is derived for the integrated scalar curvature of the embedding.

\end{abstract}

\maketitle

\section{Introduction}

Strings moving in anti-de Sitter spacetime are interesting for many reasons.
Strings lie at the heart of the AdS/CFT correspondence \cite{Maldacena:1997re, Gubser:1998bc, Witten:1998qj}. Understanding string theory enables us to study the correspondence beyond the gravity approximation.
An open string ending on the boundary is dual to a Wilson loop in the boundary theory \cite{Maldacena:1998im, Rey:1998ik}.
Strings moving on a fixed asymptotically AdS background are among the simplest holographic non-equilibrium systems \cite{Herzog:2006gh, Liu:2006ug, Gubser:2006bz}. Finally, the string worldsheet with the induced metric can be thought of as a two-dimensional toy model for gravity.

In this paper, we consider classical strings in AdS$_3$ described by the Nambu-Goto action. The theory has the remarkable property of being integrable. It can be reduced to the two-dimensional generalized sinh-Gordon theory \cite{Pohlmeyer:1975nb, DeVega:1992xc, Jevicki:2007aa, Jevicki:2009uz, Alday:2009yn, Irrgang:2012xb}. The string embedding can be reconstructed by solving an auxiliary linear problem.

\begin{figure}[h]
\begin{center}
\includegraphics[scale=0.65]{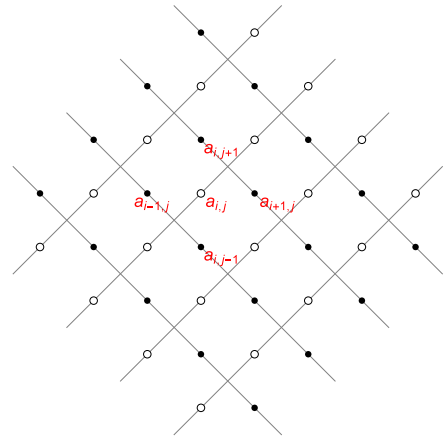}
\caption{\label{fig:sub} Kink worldlines form a rectangular lattice on the string worldsheet. The field $a_{ij}$ lives on the edges (black or white dots depending on edge orientation).
}
\end{center}
\end{figure}

The analog of a straight string in flat space is an embedding AdS$_2 \subset $ AdS$_3$ with a constant surface normal vector. More complex string solutions can be constructed by gluing AdS$_2$ patches with different normal vectors.
At the joints between adjacent segments, the string embedding contains  kinks that move with the speed of light. On the worldsheet, their worldlines form a quad lattice as seen in FIG. \ref{fig:sub}. Each square in the figure is an AdS$_2$ patch with a constant normal vector. The kink collision events will be called kink vertices.
Note that in the sinh-Gordon picture segmented strings are special solutions, because the generalized sinh-Gordon equation degenerates into the Liouville equation  \cite{Vegh:2015ska}.

In \cite{Vegh:2015ska, Callebaut:2015fsa}, the basic motion of segmented strings has been analyzed. The technique is ideally suited for numerical simulations, because the discretization is exact. This means that there are no numerical errors that otherwise may accumulate over time \cite{Vegh:2015yua}.

The present paper computes various properties of the string, including its area, energy, and scalar curvature. The area can be expressed in terms of  left (or right) variables $a_{ij}$ (or $\tilde a_{ij}$)  where $i$ and $j$ label the edges of the kink lattice. We argue that {\it classical string motion in AdS$_3$ satisfies the following equation of motion of a discrete-time relativistic Toda-type lattice}
\be
  \nonumber
  \hskip -0.15cm {1\ov a_{ij} - a_{i,j+1}}+   {1\ov a_{ij} - a_{i,j-1}} =
  {1\ov a_{ij} - a_{i+1,j}}+   {1\ov a_{ij} - a_{i-1,j}}
\ee
This is the main result of the paper. The field $a_{ij}$ is
in some sense holographic. As we will see, its value is related to the (retarded or advanced) \poincare  time when the kink corresponding to the $(ij)$ edge would reach the AdS boundary if there were no other kinks on the worldsheet.
Furthermore, we show how segmented string embeddings are obtained (up to a global $SL(2)$ transformation) from certain ``positive'' solutions of the Toda-type theory.

In the next section, we discuss the basics of segmented strings in AdS$_3$. Section III computes the area of a single patch that is bounded by four kink lines. Section IV computes the total area and the new equations of motion. Finally, reconstruction of the string from the Toda soluton is discussed. In the appendix, the string energy on the \poincare patch and the scalar curvature of the worldsheet are computed.

\section{Segmented strings}

Let us recall that the (universal cover of the) surface
\be
  \label{eq:surface}
  \vec Y \cdot \vec Y \equiv -Y_{-1}^2 - Y_0^2 + Y_1^2 + Y_2^2 = -L^2
\ee
gives an embedding of AdS$_3$ into $\RR^{2,2}$.
$L$ denotes the AdS$_3$ radius that we henceforth set to one.
A function $\vec Y(z,\bar z)$ maps the string worldsheet into this target space. The equations of motion supplemented by the Virasoro constraints are\footnote{
In the $SL(2)$ WZW model, spacetime points are $g\in SL(2, \RR)$ group elements. Classical solutions are given by $g = g_1(z)g_2(\bar z)$.
In this paper, we consider an ordinary sigma model and set the NS-NS fields (other than the metric) to zero.
Thus, classical solutions will not have such a simple product structure.
}
\bea
  \label{eq:eoms}
  \p \bar\p \vec Y - (\p \vec Y \cdot \bar\p \vec Y ) \vec Y = 0 \\
  \nonumber
  \p \vec Y \cdot \p \vec Y = \bar\p \vec Y \cdot \bar\p \vec Y = 0
\eea
where the scalar product is again that of $\RR^{2,2}$.
A normal vector to the string can be defined by
\be
  \nonumber
  N_a = { \epsilon_{abcd} Y^b \p Y^c \bar\p Y^d \over \p \vec Y \cdot \bar\p \vec Y}
\ee
It satisfies $\vec N \cdot \vec Y =\vec N \cdot \p \vec Y = \vec N \cdot \bar\p \vec Y  = 0 $ and $\vec N \cdot \vec N = 1$.
The simplest solution of (\ref{eq:eoms}) has a constant normal vector. It is the AdS$_3$ analog of an infinitely long straight string in flat spacetime.

Segmented strings are obtained by gluing worldsheet patches that have constant normal vectors \cite{Vegh:2015ska, Callebaut:2015fsa}. At the edges of the patches the string ``breaks'': on a fixed timeslice the embedding contains a kink that moves with the speed of light.

\begin{figure}[h]
\begin{center}
\includegraphics[scale=0.25]{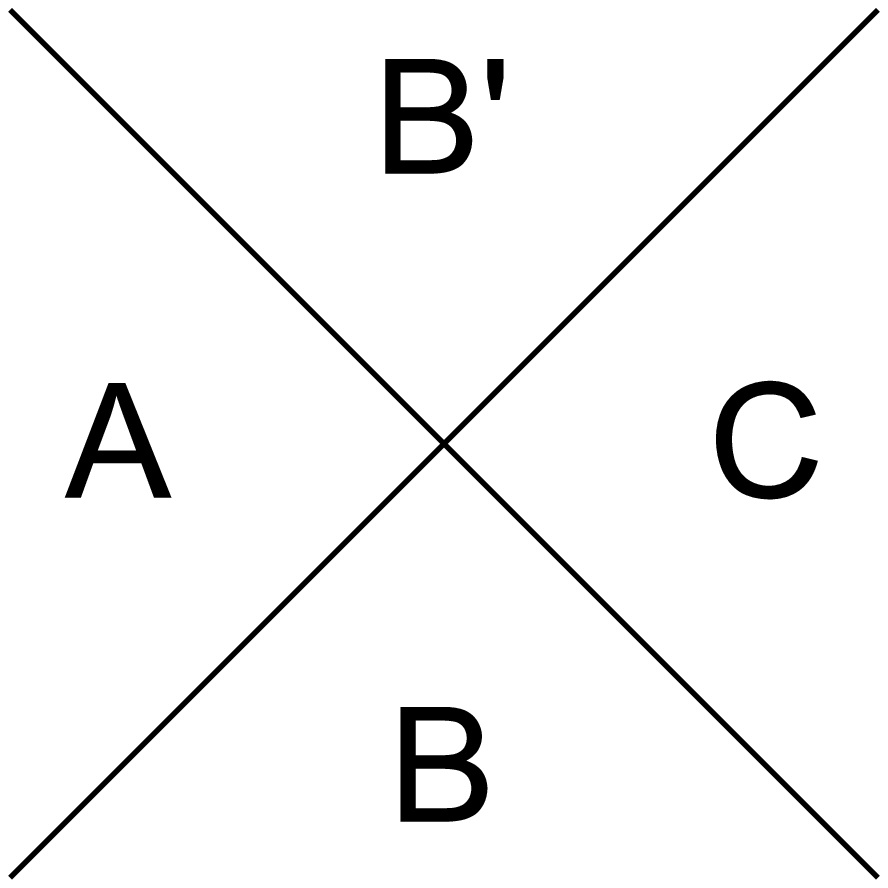}
\qquad \qquad \includegraphics[scale=0.25]{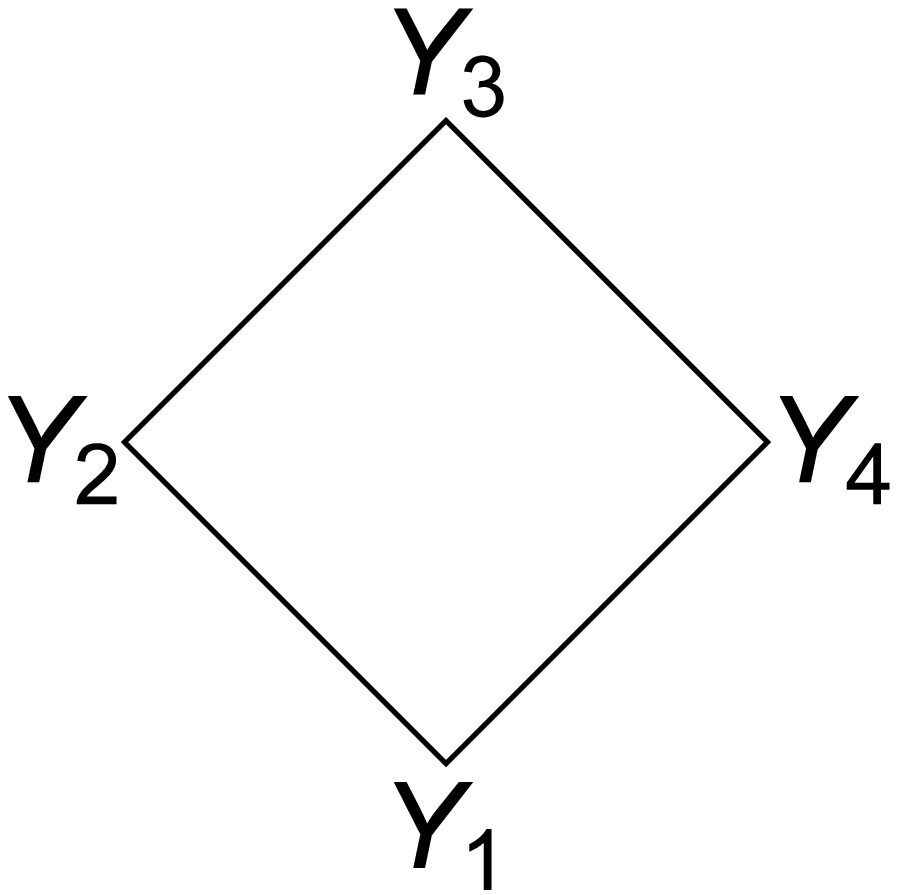}
\caption{\label{fig:latticeexpl} {\it Left:} Four AdS$_2$ patches on the worldsheet around a kink vertex. The lines are lightlike worldlines of two colliding kinks. The constant normal vectors of the for regions are $\vec A$, $\vec B$, $\vec B'$, and $\vec C$. The collision formula computes any one of these vectors from  the other three. {\it Right:} A single AdS$_2$ patch and its four vertices. The dual collision formula calculates any one of these vertices from  the other three.
}
\end{center}
\end{figure}

Normal vectors change whenever two kinks collide.
The collision on the worldsheet is shown in FIG. \ref{fig:latticeexpl}. Worldsheet time increases towards the top. The kink worldlines are indicated by two intersecting lines. Before the collision, the string consists of three segments $A,B, C$ that are characterized by three normal vectors: $\vec A$, $\vec B$, $\vec C$. We require $ \vec A \cdot \vec B = \vec B\cdot \vec C = 1$. This ensures that the kinks move on null geodesics.

After the collision, the new normal vector for the middle string piece is given by the collision formula \cite{Vegh:2015ska, Callebaut:2015fsa}
\be
  \label{eq:reflection}
  \vec B' = -\vec B +   4 {\vec A + \vec C \over (\vec A + \vec C)^2 }
\ee
One can check that $ \vec A \cdot \vec B' = \vec B' \cdot \vec C = 1$. This means that after the collision the kinks still move with the speed of light in AdS$_3$, only the directions change.

Note that $\vec A + \vec C \propto \vec B + \vec B'$. The collision formula computes any one of the four vectors from the other three by an appropriate relabeling.
Further collisions between other pairs of kinks can be computed by repeated applications of the formula.

\subsection{Dual description}

There is an  internal $SO(2,2)$ symmetry that acts on the variables (see \cite{Alday:2009yn})
\be
  \nonumber
   q_1=\vec Y, \,\,\, q_2=e^{-\alpha}\bar\p\vec Y, \,\,\, q_3=e^{-\alpha}\p\vec Y, \,\,\, q_4=\vec N
\ee
Here $\vec Y \in \RR^{2,2}$ is a point in target space, $\vec N$ is the normal vector, and the sinh-Gordon field is defined by
\be
  \nonumber
  e^{2\alpha} = \half \p \vec Y \cdot \bar\p \vec Y
\ee
The symmetry treats spacetime points and normal vectors on the same footing. Therefore, we expect to have a dual description of segmented strings in terms of points in target space instead of normal vectors.
Without proof we present here the evolution equation directly in terms of the kink vertices
\be
  \nonumber
  \vec Y_3 = -\vec Y_1  -   4 {\vec Y_2 + \vec Y_4 \over (\vec Y_2 + \vec Y_4)^2 }
\ee
Here $Y_i$ are the four vertices of a single patch as in the right of FIG. \ref{fig:latticeexpl}.
This equation is dual to (\ref{eq:reflection}). Note the sign change in the equation.
Products of vertices that are connected by a kink line are constrained, similarly to adjacent normal vectors. For instance,
\be
  \nonumber
  \vec Y_1 \cdot \vec Y_2 = -1 .
\ee
This ensures that the kink vertices $\vec Y_1$ and  $\vec Y_2$ are connected by a null vector in $\RR^{2,2}$.

\clearpage

\begin{figure}[h]
\begin{center}
\includegraphics[scale=0.6]{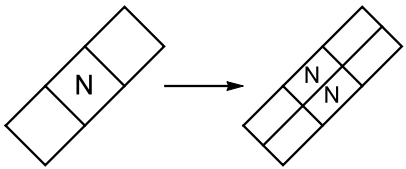}
\caption{\label{fig:sub1} Cutting patches in half by adding a zero kink. The patches inherit the normal vectors. For instance, in the new lattice two patches will have the same normal vector $\vec N$.
}
\end{center}
\end{figure}

\begin{figure}[h]
\begin{center}
\includegraphics[scale=0.6]{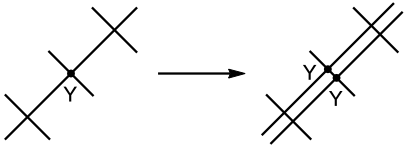}
\caption{\label{fig:sub2} Splitting a kink into two. In the new lattice, neighboring vertices are mapped to the same point in the target space. For instance, two dots will be mapped to the same $\vec Y \in \RR^{2,2}$ point in target space.
}
\end{center}
\end{figure}

\subsection{Equivalent descriptions}

A segmented string solution can be given by assigning normal vectors to faces in a square lattice. (In the dual picture, position vectors are assigned to the vertices.) The map is not one-to-one. In fact, a physical string embedding can be described by different vector lattices. There are two basic operations that preserve the physical string, but modify the lattice of vectors:
\begin{itemize}
\item {\it Adding zero strength kinks.} One can always add an extra kink line to the lattice, see FIG. \ref{fig:sub1}. This cuts a series of patches in half. On the other hand, the string embedding remains the same if the kink strength is zero: the string segment will not break at the location of the would-be kink.

\item {\it Splitting kinks.} This operation replaces a kink by a ``composite'' kink, see FIG. \ref{fig:sub2}. The new patches in between have zero area and thus the string embedding is still the same.
\end{itemize}
The two operations are dual under the $SL(2)$ transformation of the previous section. At the graphical level this can be seen by placing the dual vertices in the middle of the faces and rotating the edges by $90^\circ$.

Smooth string solutions can be obtained by considering a continuum limit with weaker and weaker kinks.
Even though segmented strings have no diffeomorphism or Weyl symmetries,
the redundancies discussed above will form the basis of the worldsheet conformal symmetry.

\begin{figure}[h]
\begin{center}
\includegraphics[scale=0.5]{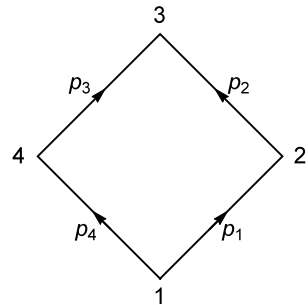}
\caption{\label{fig:momcon} A single AdS$_2$ patch of the worldsheet. The four edges are the kink worldlines where the normal vector jumps. In $\RR^{2,2}$ these are straight null lines with direction vectors $\vec p_i$.
}
\end{center}
\end{figure}

\section{Area of a single patch}

What is the string area in terms of the discrete data that defines segmented strings? Let us first focus on a single patch with a constant normal vector, see FIG. \ref{fig:momcon}. The boundary of the worldsheet patch consists of four kink lines. In the target space, these are mapped to straight null lines (a consequence of the Virasoro constraints).
Let us denote the four vertices of the patch by $\vec V_i  \in \RR^{2,2}$. We have $(\vec V_i)^2 = -1$.
Let us define the lightlike direction vectors as in the figure
\bea
  \nonumber
  \vec p_1 = \vec V_2 - \vec V_1 &\qquad &
  \vec p_2 = \vec V_3 - \vec V_2 \\
  \label{eq:difvec}
  \vec p_3 = \vec V_3 - \vec V_4 &\qquad &
  \vec p_4 = \vec V_4 - \vec V_1
\eea
We have
\be
  \nonumber
  (\vec p_i)^2 = 0  \qquad \textrm{and} \qquad  \vec p_1 + \vec p_2  = \vec p_3 + \vec p_4
\ee
The latter equation can be interpreted as ``momentum conservation'' during the scattering of two massless scalar particles with initial and final momenta $\vec p_{1,2}$ and $\vec p_{3,4}$, respectively. The area of the patch is analogous to a scattering amplitude that is invariant under the $SO(2,2)$ isometry group of AdS$_3$.  The only independent invariants are the Mandelstam variables $s = (\vec p_1+ \vec p_2)^2$ and  $u= (\vec p_1 - \vec p_4)^2$. The patch area is then
\be
  \nonumber
  A_\textrm{patch} = L^2 \mathcal{F}\le( {u \over s} \ri)
\ee
where $L$ is the AdS$_3$ radius (henceforth set to one) and $\mathcal{F}(x)$ is a dimensionless function to be determined.

Let us consider an AdS$_2 \subset$ AdS$_3$ patch with normal vector $N = (0,0,0,1)$. Points on the surface are of the form $X = (x_{-1}, x_0, x_1, 0)$ with $x_{-1}^2+x_0^2 -x_1^2= 1$. Let us fix a parameter $a \in (0,1)$ and consider four points
\bea
  \nonumber
  \vec V_1 &=& (a,  -\sqrt{1-a^2}, \quad \,\,\,\, 0, \quad \,\,\,\,\, 0) \\
  \nonumber
  \vec V_2 &=& (a^{-1}, \quad 0, \, \sqrt{-1+a^{-2}}, \,\,\,\, 0) \\
  \nonumber
  \vec V_3 &=& (a, \, \,\,\, \sqrt{1-a^2}, \qquad 0, \quad \,\,\,\, 0) \\
  \nonumber
  \vec V_4 &=& (a^{-1},  \quad 0, \, -\sqrt{-1+a^{-2}},   0)
\eea

These points satisfy $(\vec V_i)^2 = -1$.
It is easy to check that the corresponding difference vectors from (\ref{eq:difvec}) indeed
satisfy $(\vec p_i)^2 = 0$. Cusps on the patch boundaries move along these lightlike vectors.
Let us now compute the area of this patch. Define
\be
  \nonumber
  \vec X_0(\tau, \sigma) = [(1-\sigma) \vec V_1 + \sigma \vec V_2] (1-\tau) + \tau \vec V_3
\ee
and $X = X_0 / |X_0|$. Thus, $X(\tau, \sigma)$  parametrizes half of the patch if $\sigma, \tau \in (0,1)$.
After a lengthy calculation, the induced metric $g$ gives
\be
  \nonumber
  \sqrt{-g} =  {2(1-a^2)(1-\tau) \over \le[  1 -4(1-a^2)(1-\sigma)(1-\tau)\tau \ri]^{3\ov 2} }
\ee
Integrating with respect to $\tau$ and $\sigma$ gives the area of half of the patch. From this we get
\be
  \nonumber
  A_\textrm{patch} = -4\log a
\ee
For our patch, the ratio of the Mandelstam variables is given by $s/u = -a^2$.
Combining these results fixes $\mathcal{F}(x)$ and we get the covariant formula
\be
  \label{eq:covarea}
  A_\textrm{patch} = \log\le[{(\vec p_1 - \vec p_4)^2 \over (\vec p_1 + \vec p_2)^2 } \ri]^2
\ee
Positiveness of the area requires
\be
  \nonumber
  |\vec p_1 - \vec p_4| > |\vec p_1 + \vec p_2|
\ee
This constrains the possible values of $\vec p_i$.

\subsection{Helicity spinors}

In the spinor helicity formalism, one exhibits lightlike momentum vectors as products of spinors.
We define
\bea
  \nonumber
  \sigma^\mu &=& ( 1, -i \sigma_2, \sigma_1, \sigma_3 ) \\
  \nonumber
   p_{a\dot a} &=& \sigma^\mu_{a\dot a} p_\mu
\eea
Since $p^2 = \det(p_{a\dot a}) = 0$, we can write
\be
  \nonumber
  p_{a\dot a} = \lambda_a \tilde\lambda_{\dot a}
\ee
Since $\lambda_a \to s \lambda_a$, $\tilde\lambda_{\dot a} \to {1\ov s} \tilde\lambda_{\dot a}$ does not change $p_{a\dot a}$, there is a new gauge redundancy in this description.
The spinors can be chosen to be real in $\RR^{2,2}$.

\begin{figure}[h]
\begin{center}
\includegraphics[scale=0.6]{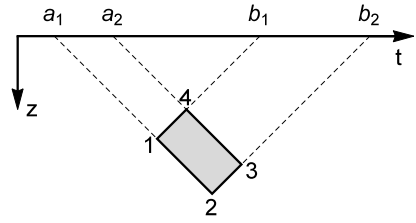}
\caption{\label{fig:pp} Computing the area of a single rectangular patch of AdS$_2$. The figure shows the \poincare patch. The $t$-axis on the top is the boundary. The string patch (gray rectangle) can be specified by four points on the boundary: $a_{1,2}$ \& $b_{1,2}$. The four vertices of the rectangle are connected to these points by the null lines (dashed).
}
\end{center}
\end{figure}

Let us now define the $SO(2,2)$ invariants,
\bea
  \nonumber
  \langle\lambda_i, \lambda_j \rangle &=& \epsilon_{ab} \lambda_1^a \lambda_2^b \\
  \nonumber
  \[ \tilde\lambda_i, \tilde\lambda_j \] &=& \epsilon_{\dot a \dot b} \tilde\lambda_1^{\dot a} \tilde\lambda_2^{\dot b}
\eea
and consider two vectors $\vec p$ and $\vec q$ with the  decomposition
\bea
  \nonumber
  p_{a\dot a} &=& \lambda_a \tilde \lambda_{\dot a} \\
  \nonumber
  q_{a\dot a} &=& \kappa_a \tilde \kappa_{\dot a}
\eea
Their product is expressed as
\be
  \nonumber
  \vec p \cdot \vec q = \langle \lambda, \kappa \rangle [\tilde\lambda, \tilde\kappa]
\ee
This allows us to write (\ref{eq:covarea}) in the form
\be
  \nonumber
  A_\textrm{patch} = \log\le({\langle\lambda_1, \lambda_4 \rangle [\tilde\lambda_1, \tilde\lambda_4 ]
  \over \langle\lambda_1, \lambda_2 \rangle [\tilde\lambda_1, \tilde\lambda_2 ] } \ri)^2
\ee
``Momentum conservation'' can be written as
\be
  \nonumber
  \sum_{i=1}^4 \lambda_i^a \tilde\lambda_i^{\dot a}=0
\ee
where $i$ runs over the four edges around the patch.
This can be used to cast the area formula in the form
\be
  \label{eq:arealeft}
   A_\textrm{patch} = 2 \log\le|  {\langle\lambda_1, \lambda_4 \rangle\langle\lambda_2, \lambda_3 \rangle
  \over \langle\lambda_1, \lambda_2 \rangle\langle\lambda_3, \lambda_4 \rangle  } \ri|
\ee
Note that this expression contains only ``left-handed'' variables. There is a similar formula with only  ``right-handed'' $\tilde\lambda$ spinors. Let us stress that the spinors cannot take on arbitrary values because the area must be non-negative.

\subsection{Global variables}

Clearly, formula (\ref{eq:arealeft}) does not depend on the modulus of $\lambda_i$, i.e. it is gauge-invariant. By defining $|\lambda_i| e^{i\alpha_i} := \lambda_i^1 + i\lambda_i^2 $  we get
\be
  \label{eq:patcharea}
  A_\textrm{patch} =  2\log\le|   {\sin(\alpha_{1} - \alpha_{4})\sin(\alpha_{2} - \alpha_{3}) \over \sin(\alpha_{1} - \alpha_{2})\sin(\alpha_3 - \alpha_{4})} \ri|
\ee
For a given AdS$_2$ with a fixed normal vector, the four angles fully specify the four (infinite) straight kink lines in $\RR^{2,2}$.
Changing an angle means that we move the corresponding line on AdS$_2$. Note that the area diverges whenever two adjacent angles are equal. This corresponds to a configuration where a kink collision takes place on the UV boundary of AdS.

\begin{figure}[h]
\begin{center}
\includegraphics[scale=0.5]{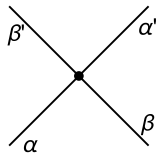}
\caption{\label{fig:aabb} Two kinks crossing. The (left) angles change from $\alpha, \beta$ to $\alpha', \beta'$. If the collision point is specified in $\RR^{2,2}$, then the angles determine the four kink vectors via (\ref{eq:fourvel}).
}
\end{center}
\end{figure}

When two kinks cross each other, the angles generically change, see FIG. \ref{fig:aabb}. In a special case, e.g. $\alpha = \alpha'$, the $\beta$ and $\beta'$ lines are ``zero kinks'': they can be removed from the kink lattice without changing the string embedding. Such redundancies have been discussed in section II.

In order to have a better understanding of the $\alpha$ angles, let us compute them in \poincare coordinates for a patch on a particular AdS$_2 \subset$ AdS$_3$ whose normal vector is $\vec N = (0,0,0,1)$.
Using the \poincare parametrization of AdS$_3$
\be
  \nonumber
  \vec Y = \le( {t^2 -z^2 - x^2 -1 \ov 2z}, \, {t\ov z} ,  \, {t^2 -z^2 - x^2 +1 \ov 2z},  \,{x\ov z}\ri)
\ee
and setting $x=0$, we obtain a parametrization of our AdS$_2$.
The induced metric is
\be
  \nonumber
  ds^2 = {-dt^2 + dz^2 \ov z^2}
\ee
The patch in AdS$_2$ is bounded by $\pm 45^\circ$ lines on the $t-z$ plane. This is shown in FIG. \ref{fig:pp}. The lines intersect the AdS$_2$ boundary   at $t = a_1, a_2, b_1, b_2$ as in the figure.
The four vertices are $V_1 = v_{11}$,  $V_2 = v_{12}$,  $V_3 = v_{22}$,  $V_4 = v_{21}$, where
\be
  \nonumber
  v_{ij} = \le({1-a_i b_j \ov a_i - b_j},  \,  {a_i + b_j \ov b_j - a_i},  \, {1+a_i b_j \ov  b_j - a_i} , \, 0 \ri) .
\ee
From this, the lightlike boundary vectors $p_i$ can be computed. The corresponding left angles $\alpha_i$ are (after a ${\pi \ov 4}$ shift in order to have simpler expressions)
\bea
  \nonumber
  \tan \alpha_1 =  a_1 
  &\qquad &
  \tan \alpha_2 =  b_2 
  \\
  \tan \alpha_3 =  a_2 
  &\qquad &
  \tan \alpha_4 =  b_1 
  \label{eq:globallocal}
\eea
The patch area is given by the formula (\ref{eq:covarea}) that yields
\be
  \nonumber
  A_\textrm{patch} = 2\log\le| {(a_1-b_1)(a_2-b_2) \ov (a_2-b_1)(a_1-b_2) }\ri|
\ee
This expression is equal to (\ref{eq:patcharea}).

\begin{figure}[h]
\begin{center}
\includegraphics[scale=0.55]{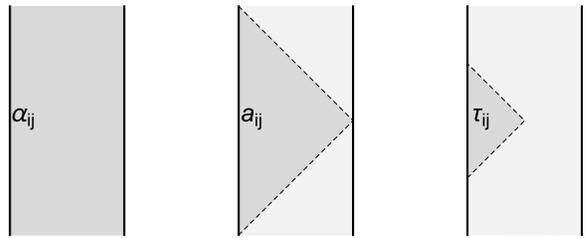}
\caption{\label{fig:coo} The three boundary time coordinates can parametrize different (shaded) regions in AdS$_2$: (i) global coordinate $\alpha_{ij}$, (ii) \poincare coordinate $a_{ij}$, and (iii) Schwarzschild coordinate $\tau_{ij}$.
}
\end{center}
\end{figure}

\subsection{\poincare and Schwarzschild variables}

Although the patch normal vectors are generically different from $(0,0,0,1)$, motivated by (\ref{eq:globallocal}) we define the {\it \poincare variables}
\be
  \label{eq:pdef}
  a_i := \tan \alpha_i
\ee

We have seen that for the particular normal vector $\vec N = (0,0,0,1)$, the \poincare variables are ``holographic coordinates'':  they correspond to  retarded and advanced times when light rays reach the boundary.
Since the relationship between global and \poincare AdS times is the same as in (\ref{eq:pdef}), the $\alpha$ variables will henceforth be called {\it global variables}.

There is a third set of variables that is easiest to see using Mikhailov's  parametrization of the AdS$_2$ subspace \cite{Mikhailov:2003er}. Let us consider (\ref{eq:mikh}) in the appendix and plug in the quark motion
\be
  x_1(t) =  \sqrt{1 + t^2}
\ee
Then, (\ref{eq:mikh}) gives an AdS$_2$ with normal vector $(0,0,1,0)$.

Let us consider a kink line on this AdS$_2$ subspace. Its location is specified by a global variable $\alpha$. Let $t_{\textrm{AdS}_3}$ denote the  \poincare AdS$_3$ time when this null geodesic intersects the boundary.
A short calculation gives
\be
  \nonumber
   t_{\textrm{AdS}_3} = \tan 2\alpha
\ee
The string endpoint on the boundary of AdS$_3$ is a quark that suffers constant acceleration.
Its proper time is related to the AdS$_3$ time coordinate as $\tau = \sinh^{-1} t$.
Thus, we have
\be
  \label{eq:schwdef}
  \tau  = \sinh^{-1}\tan 2\alpha = 2\tanh^{-1}\tan \alpha
\ee
and this equation defines the {\it Schwarzschild variables} $\tau$ in the general case.

Let us summarize the results in this section. The global $\alpha$ variables are the angles of the {\it left} helicity spinors $\lambda$. The \poincare and Schwarzschild fields are simply computed via (\ref{eq:pdef}) and (\ref{eq:schwdef}), respectively. Different variables are related to different coordinate systems on AdS$_3$. This is shown in FIG. \ref{fig:coo}.
Similarly, one defines right-handed fields starting from the spinors $\tilde\lambda$. These variables will be denoted $\tilde\alpha$, $\tilde a$, and $\tilde \tau$.
We note that the map between the left and right fields is non-local.
Finally, the area of the string can be expressed in either left or right variables, see eqn. (\ref{eq:patcharea}).
In the next section, we will compute the string action and show how the string embedding can be reconstructed from the angle variables.

\section{Total area}

The total area of the string is the sum of individual patch areas. From (\ref{eq:patcharea}), we have
\be
  \nonumber
  A = \sum_{f \in \textrm{patches}}  \log\le(   {\sin(\alpha_{f_1} - \alpha_{f_4})\sin(\alpha_{f_2} - \alpha_{f_3}) \over \sin(\alpha_{f_1} - \alpha_{f_2})\sin(\alpha_{f_3} - \alpha_{f_4})} \ri)^2
\ee
where $f_{1,2,3,4}$ label the four edges around the patch $f$ and $\alpha_i$ are the left angles.
The action can be cast in the form,
\be
  \label{eq:gaction}
  A = 2 \sum_{i,j}    \log\le| {\sin(\alpha_{i,j} - \alpha_{i+1,j}) \over \sin(\alpha_{i,j} - \alpha_{i,j+1})} \ri|
\ee
where $i$ and $j$ are coordinates on a square lattice, see FIG. \ref{fig:sub}.
There is a similar formula that involves only the right-handed angles.
The total area can be expressed in terms of  \poincare variables as well ($a_{ij} := \tan \alpha_{ij}$)
\be
  \label{eq:aaction}
  A = 2 \sum_{i,j}   \log\le| a_{i,j}  - a_{i+1,j}  \over   a_{i,j}  - a_{i,j+1} \ri|
\ee
Finally, in Schwarzschild variables ($\tanh{{\tau_{ij} \ov 2}} := \tan \alpha_{ij}$)
we have
\be
  \label{eq:raction}
  A = 2 \sum_{i,j}    \log\le|  {\tanh {\tau_{i,j}\ov 2} - \tanh {\tau_{i+1,j}\ov 2}}  \over  \tanh {\tau_{i,j}\ov 2} - \tanh {\tau_{i,j+1}\ov 2} \ri|
\ee

Note that patches are assumed to be located entirely in the bulk. Whenever a patch intersects the boundary of AdS$_3$, the area must be regularized.

\subsection{Equation of motion}

The Nambu-Goto action is equal to the area of the string which can be extremized by varying the left fields. The expression for the total area defines an action for a new Toda-type theory. Classical segmented string solutions yield solutions to this theory. 

Let us first consider the action in \poincare variables. The equation of motion is ${\delta A \ov \delta a}=0$. From (\ref{eq:aaction}) we have
\be
  \label{eq:eom}
  \hskip -0.1cm {1\ov a_{i,j} - a_{i,j+1}}+   {1\ov a_{i,j} - a_{i,j-1}} =
  {1\ov a_{i,j} - a_{i+1,j}}+   {1\ov a_{i,j} - a_{i-1,j}}
\ee
The same equation is satisfied by the $\tilde a$ variables of the right-handed theory.
The $a$ field computed from a string solution will satisfy this equation. We recognize this equation of motion as that of a {\it time-discretized relativistic Toda-type lattice with a cubic Poisson bracket}, see (10.10.6) on page 442 in \cite{suris}. Note that (\ref{eq:eom}) can also be thought of as a local version of the equation of motion of the discrete-time Caloger-Moser model \cite{Nijhoff:1994ic}.

From (\ref{eq:raction}) we have another local equation
\bea
  \nonumber
  {1\ov \tanh(\tau_{i,j} - \tau_{i,j+1})}+   {1\ov \tanh(\tau_{i,j} - \tau_{i,j-1})} = \qquad\qquad \\
  \nonumber
  {1\ov \tanh(\tau_{i,j} - \tau_{i+1,j})}+   {1\ov \tanh(\tau_{i,j} - \tau_{i-1,j})}
\eea
which is the same as (10.8.7) on page 440 in \cite{suris}.
Similar equation follows from (\ref{eq:gaction}) with $\tan(x)$  functions in the denominators.

Initial conditions can be specified by setting two rows in the lattice (e.g. $a_{i0}$ and $a_{i1}$).

A trivial solution is  given by considering a lattice of zero kinks. For such a lattice, any two angles are the same if they lie on the same kink line,
\be
  \nonumber
  a_{i,j} =  \left\{ \begin{array}{llll}
  u(i+j)  & &  & \textrm{if $i+j$ is  odd}\\
  v(i-j)  & &  & \textrm{if $i+j$ is even}
\end{array} \right.
\ee
This solution describes a single AdS$_2$ with a constant normal vector. The physical string embedding does not depend on $u$ and $v$.

\subsection{Reconstructing the embedding}

The string embedding can be reconstructed from a solution of (\ref{eq:eom}). There are some caveats, however.
\begin{itemize}

\item The symmetry group of AdS$_3$ is $SO(2,2) = SL(2)_L \times SL(2)_R$. The two $SL(2)$s act on the $\lambda_a$ left and $\tilde\lambda_{\dot a} $ right spinors, respectively. Since left fields do not change under $SL(2)_R$, $a_{ij}$ only determines the string embedding up to such global transformations. Similarly, $\tilde a_{ij}$ determines only an orbit of $SL(2)_L$.

\item
Note that there is a $\ZZ_2$ ambiguity in assigning a kink lattice to the lattice of black-and-white dots in FIG. \ref{fig:sub} if their color is unknown.
Thus, the symmetry between kink collision points in spacetime and AdS$_2$ patch normal vectors is manifest in the Toda description.
As a consequence, two different embeddings may be constructed from $a_{ij}$.

\item Not all Toda solutions correspond to string embeddings. The area of the string patches must be non-negative. This is only true for solutions satisfying
\be
  \nonumber
   (a_{i+1,j}  - a_{i,j-1})(a_{ij}  - a_{i+1,j-1}) > 0
\ee
for the four angles around any patch (i.e. ${ij}$ is a white dot in FIG. \ref{fig:sub}).

\end{itemize}

In the following we sketch the procedure for rebuilding the string solution.
Let us fix a spacetime point $\vec X \in \RR^{2,2}$ that will correspond to a particular kink collision event in FIG. \ref{fig:sub}. The four angles around the vertex are
\be
  \label{eq:angles}
  \alpha_{ij}, \,\,\,  \alpha_{i+1,j}, \,\,\, \alpha_{i,j+1}, \,\,\, \alpha_{i+1,j+1}
\ee
For any one of these angles, the corresponding kink vector is computed  to be
\be
  \label{eq:fourvel}
  \vec p \propto  \left( \begin{array}{lllll}
  -X_0 &+& X_2 \sin 2\alpha  &+& X_1 \cos 2\alpha  \\
 \,  X_{-1} &-& X_1 \sin 2\alpha  &+& X_2 \cos 2\alpha  \\
 \,\,\,\, X_2 &-& X_0 \sin 2\alpha  &+& X_{-1} \cos 2\alpha  \\
  -X_1 &+& X_{-1} \sin 2\alpha  &+& X_0 \cos 2\alpha
\end{array} \right)
\ee
for which $p^2 = 0$. Two adjacent kink vectors, e.g. $\vec p_{ij}$ and $\vec p_{i+1,j}$, define an AdS$_2$ with a constant normal vector that contains the points $\vec X + \lambda \vec p_{ij}$ and $\vec X + \lambda' \vec p_{i+1,j}$ for any $\lambda$ and  $\lambda'$. Let us pick two adjacent angles $\alpha$ and $\beta$ from (\ref{eq:angles}). The corresponding kink vectors span an AdS$_2$ patch with normal vector
\bea
  \nonumber
   &\vec N(\alpha,\beta) = {1\ov \sin(\alpha-\beta)} \times \hskip 6cm  &  \\
  \nonumber
  &
  \hskip -0.8cm  \times \left( \begin{array}{l}
   \, \, \, \, \, \, \, \, X_0 \cos(\alpha-\beta) - X_1 \cos(\alpha+\beta)  - X_2 \sin(\alpha+\beta)  \\
   -X_{-1} \cos(\alpha-\beta) - X_2 \cos(\alpha+\beta)  + X_1 \sin(\alpha+\beta)  \\
   -X_2 \cos(\alpha-\beta) - X_{-1} \cos(\alpha+\beta)  + X_0 \sin(\alpha+\beta)  \\
    \, \, \, \, \, X_1 \cos(\alpha-\beta) - X_0 \cos(\alpha+\beta)  - X_{-1} \sin(\alpha+\beta)
\end{array} \right)
   &
\eea
We have seen that once the location of a kink vertex is fixed in spacetime, the four angles around it fully specify the four kink vectors. Similarly, if a normal vector is known, four angles fully specify the four boundaries of the corresponding patch. The boundary edges then intersect each other at new kink vertices and the kink vectors around those can also be computed. This procedure can be repeated until the entire worldsheet embedding is recovered.

\section{Discussion}

Let us consider space-like string embeddings in anti-de Sitter spacetime. A smooth open string that ends on a curve $\mathcal{C}$ in the boundary can be approximated by another string that ends on a zigzag line in the boundary whose segments are lightlike and which itself is sufficiently close to $\mathcal{C}$ \cite{Alday:2009yn}\footnote{This is probably best visualized by imagining a soap bubble (minimal surface) stretching on a zigzag wire.}.
In the case of Lorentzian embeddings, a Lorentzian zigzag worldline constitutes a singular limit, because the boundary quark sitting on the endpoint of the string would radiate off an infinite amount of energy at the turning points. If the quark velocity cannot jump, how can smooth strings be approximated by strings that are described by discrete data? A solution is provided by segmented strings \cite{Vegh:2015ska, Callebaut:2015fsa}. In this case only the  acceleration of the quark jumps whenever kinks enter or leave the string.
Kinks between the segments move with the speed of light and (between collisions) their velocities are constant vectors in the embedding $\RR^{2,2}$ spacetime.
When kinks collide, the new normal vector to the string is given by the collision formula (\ref{eq:reflection}).

In this paper, we have computed the area of segmented strings in terms of cross-ratios of helicity spinors. These spinors arise from the decomposition of the kink vectors. The string area equals the Nambu-Goto action which we have expressed purely in terms of left (or right) angle variables.
We have argued that the time evolution of the segmented string can be described by the evolution equation of a discrete-time Toda-type lattice.

The Toda-type lattice contains only left- or right-handed fields that are exchanged by a parity transformation ($X_2 \to -X_2$ in $\RR^{2,2}$).
On the other hand, the sigma model has manifest parity symmetry and is probably best written as a constrained sum of a left and a right Toda-type lattice.
For each AdS$_2$ patch on the worldsheet, there is a ``momentum constraint'' that involves spinors of both handedness
\be
  \label{eq:momconx}
  \sum_{i=1}^4 \lambda_i^a \tilde\lambda_i^{\dot a} = 0
\ee
The meaning of this equation is that the boundary of a patch is a closed loop in spacetime.
(There is a similar constraint for every kink collision vertex as well.) 
The left and right variables are therefore not independent: they are ``classically entangled''. It would be interesting to relate the Toda-type lattice to a matrix model perhaps via a (relativistic) Calogero-Moser theory.

\vskip 0.5cm \quad

\vspace{0.2in}   \centerline{\bf{Acknowledgments}} \vspace{0.2in}
This work was supported by the Center of Mathematical Sciences and Applications at Harvard University.
The author would like to thank Daniel Harlow and Burkhard Schwab for discussions and Douglas Stanford for comments on the manuscript.

\section*{Appendix}

\subsection{String energy}

In this section, we compute the energy of the string on the \poincare patch of AdS$_3$. The metric is
\be
  \nonumber
  ds^2 = {-dt^2 + dx^2 + dz^2 \over z^2}
\ee
Let us consider the action
\be
  \nonumber
  S =  -{1 \ov 4\pi \alpha'} \int d^2 \sigma \sqrt{-h} h^{ab} \p_a X^\mu\p_b X^\nu G_{\mu\nu}
\ee
where $X^\mu$ are embedding coordinates,  $h_{ab}$ is the worldsheet metric, and $a,b \in \{\tau, \sigma \}$.
One defines the worldsheet currents of target space energy-momentum
\be
  \nonumber
  P^a_\mu = -{1 \ov 2\pi \alpha'} \sqrt{-h}\,h^{ab} G_{\mu\nu} \p_b X^\nu
\ee
From the equation of motion it follows that
\be
  \nonumber
  \p_a P^a_\mu - \Gamma^\kappa_{\mu\lambda} \p_a X^\lambda P^a_\kappa = 0
\ee
Defining $p^a_\mu = { P^a_\mu \over \sqrt{-h}}$, and substituting the induced metric $g_{ab}=\p_a X^\mu\p_b X^\nu G_{\mu\nu}$ for $h_{ab}$, this can be written as
\be
  \nonumber
  \nabla_a p^a_\mu - \Gamma^\kappa_{\mu\lambda} \p_a X^\lambda p^a_\kappa = 0
\ee

\begin{figure}[h]
\begin{center}
\includegraphics[scale=0.7]{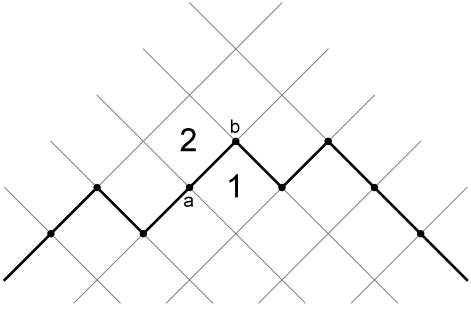}
\caption{\label{fig:12} Integration of spacetime currents along lightlike patch boundaries on the worldsheet. Kinks are indicated by thin lines. The thick line shows a possible path of integration. Contributions to the energy corresponding to elementary lightlike segments can be computed (e.g. $E_{1,2}$ for the segment between $a$ and $b$) and the sum gives the total energy of the string.
}
\end{center}
\end{figure}

\noindent
where $\nabla$ is the covariant derivative with respect to $g_{ab}$. Note that the target space index $\mu$ is only a spectator when the derivative is taken. The second term is
\be
  \nonumber
  \Gamma^\kappa_{\mu\lambda} \p_a X^\lambda p^a_\kappa = \half G_{\nu\lambda, \mu} \p_a X^\lambda\p_b X^\nu g^{ab}
\ee

If $G_{\nu\lambda}$ is independent of $X^\mu$ for some $\mu$, then $\zeta^\alpha = \delta^\alpha_\mu$ is a Killing vector. Then $ \zeta^\alpha \nabla_a p^a_\alpha =0$ and one can define the conserved quantity
\be
  \nonumber
  E_\zeta = - \int d\sigma   \zeta^\alpha P^\tau_\alpha
\ee
that satisfies $\p_\tau E_\zeta = 0$. We are going to use $\zeta = \p_t$ in the following.

Energy expressions are typically complicated (see \cite{Callebaut:2015fsa}). In order to simplify the results, we perform the integration on the worldsheet along a path that consists of lightlike patch boundaries (instead of  constant $\tau$ slices).

In terms of the \poincare coordinates, a single AdS$_2$ patch is a contracting and expanding semi-circle. Let $x_1$ denote the path of the quark in the boundary
\be
  \nonumber
  x_1(t) = X_1 + \sqrt{R_1^2 + (t-T_1)^2}
\ee
This is a hyperbola, parametrized by $X_1, T_1$, and $R_1$. The subscript indicates the patch, see FIG. \ref{fig:12}.
Using Mikhailov's result, the string embedding is given by
\bea
  \nonumber
  t(\tau,z) &=& \tau + {z\ov \sqrt{1-x_1 '(\tau)^2}} \\
  \nonumber
  z(\tau,z) &=& z \\
  \label{eq:mikh}
  x(\tau,z) &=&  x_1(\tau) + {z x_1'(\tau)\ov \sqrt{1-x_1'(\tau)^2}}
\eea
Here $\tau$ plays the role of retarded time.
The relationship between the normal vector $\vec N$ and the parameters  of the hyperbola are
\be
  \nonumber
  (T_1, \, X_1, \, R_1) = \le(  {-N_{0} \over N_{-1} + N_2}, \ {-N_{1} \over N_{-1} + N_2}, \ {1 \over | N_{-1} + N_2 |} \ri) .
\ee
The induced metric on the worldsheet is
\be
\nonumber
 g = {1 \over z^2 \sqrt{R_1^2 + (\tau - T_1)^2} }\left(
\begin{array}{cc}
 {z^2 - R_1^2 \over \sqrt{R_1^2 + (\tau - T_1)^2}} & -R_1 \\
  -R_1 & 0
\end{array}
\right)
\ee
\bwt
The current takes the form
\be
  \nonumber
{p^a}{}_\mu(\tau, z) = {1 \ov 2\pi \alpha'}\left(
\begin{array}{ccc}
 -\frac{R_1^2+(\tau -T_1)^2}{R_1^2} & \frac{\sqrt{R_1^2+(\tau -T_1)^2}}{R_1} & \frac{\sqrt{R_1^2+(\tau -T_1 )^2} (\tau -T_1)}{R_1^2} \\
 -\frac{z \left(R_1 (\tau -T_1)+z \sqrt{R_1^2+(\tau -T_1)^2}\right)}{R_1^3} & \frac{z^2}{R_1^2}-1 & \frac{z \left( z(\tau -T_1)+R_1 \sqrt{R_1^2+(\tau -T_1 )^2}\right)}{R_1^3} \\
\end{array}
\right)
\ee
\ewt
We want to integrate $P^\tau_t$ along the lightlike boundary between the patch $(T_1, X_1, R_1)$ and another patch $(T_2, X_2, R_2)$. This translates to integrating over $z$ at a fixed $\tau$.  The interpretation of $\tau$ is that the kink between the two patches reaches the boundary at \poincare time $t = \tau$ (unless it collides with other kinks).

The value of $\tau$ can be computed by requiring that at this time the quark velocity is continuous
\be
  \nonumber
  \p_\tau x_1(\tau, 0) =   \p_\tau x_2(\tau, 0)
\ee
from which
\be
  \nonumber
  \tau = {R_2 T_1 + R_1 T_2 \over R_1 + R_2}
\ee
Finally, the contribution of the kink to the total energy is given by the integral
\be
  \nonumber
  E_{1,2} = \int_{z_a}^{z_b} dz \sqrt{-g} \, p^\tau_t = {{1 \ov z_b} - {1 \ov z_a}  \ov 2\pi \alpha' } \sqrt{1+\le({T_1 - T_2 \ov R_1 - R_2}\ri)^2 }
\ee
where $z_a$ and $z_b$ are the $z$ coordinates of the points where the kink is created and annihilated, respectively. Note that the formula is symmetric under $1 \leftrightarrow 2$ as it should be. Furthermore, $X_{1}$ and $X_{2}$ have dropped out.
Note that for the correct null cusp limit one takes  $T_1 \to T_2$ before $R_1 \to R_2$.
The total energy of the string is given by the sum of all $E_{i,j}$ along a zigzag path on the patch boundaries, see FIG. \ref{fig:12}.

\clearpage

\begin{figure}[h]
\begin{center}
\includegraphics[scale=0.7]{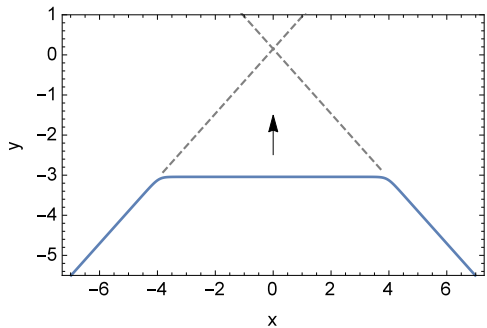}
\caption{\label{fig:smoothcoll} Collision of two smooth kinks in Minkowski spacetime. The left and right segments are static semi-infinite lines. The middle segment moves with a constant velocity. The velocity changes sign at the collision.
}
\end{center}
\end{figure}

\subsection{Scalar curvature}

Let us consider a static string that hangs from the boundary of AdS$_3$. The induced metric on the worldsheet is that of AdS$_2$ and the Ricci scalar is $R=-2$. This is the only intrinsic curvature invariant that one can compute in two dimensions.
If the string endpoint is perturbed, kinks (or waves in the continuum case) will travel down the string. However, the scalar curvature does not change.  The reason for this is simple and best understood through a flat space analogy. Consider, for instance, a cube and cut out the top and bottom faces.

What's left is the four adjacent side faces that can be unwrapped and arranged on a plane. The four edges are the analogs of the kinks that move in the same direction.

The induced metric on the four faces is clearly flat.
Only when kinks {\it collide} can the curvature differ from the constant value.
In this section, we compute the integrated Ricci scalar at collision points.

Since kink collisions happen at single points, the background curvature can be neglected. Thus, one can analyze the problem in 2+1 dimensional Minkowski space with coordinates $x^{0,1,2}$. In order to handle the divergence in the curvature at the collision point, we smoothen the step functions corresponding to kinks.

A string with two smooth kinks colliding on it is given by the embedding
\bea
  \nonumber
  x^0 &=& {(2+A^2)(\sigma^+ + \sigma^-) \over 2 \sqrt{2}} - {A^2 (\tanh \epsilon \sigma^- + \tanh \epsilon \sigma^+ ) \over 2 \sqrt{2} \epsilon } \\
  \nonumber
  x^1 &=& {(2-A^2)(\sigma^+ - \sigma^-) \over 2 \sqrt{2}} - {A^2 (\tanh \epsilon \sigma^- - \tanh \epsilon \sigma^+ ) \over 2 \sqrt{2} \epsilon } \\
  \nonumber
  x^2 &=&  -{A ( \log \cosh \epsilon \sigma^- +\log \cosh \epsilon \sigma^+ ) \over \epsilon }
\eea
Here $\sigma^+$ and $\sigma^-$ are lightcone coordinates on the worldsheet. $A$ parametrizes the kink strengths and $\epsilon$ is related to the smoothness of the step functions.

\begin{figure}[h]
\begin{center}
\includegraphics[scale=0.45]{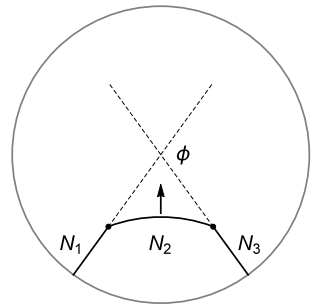}
\caption{\label{fig:angle} Collision of two kinks (dots in the figure) in global AdS$_3$. The figure shows a constant time slice before the collision. Patches $N_1$ and $N_3$ are static straight lines in these coordinates. The angle between them is denoted $\phi$. The length of the $N_2$ piece decreases and vanishes at the time of collision. After the collision, its velocity is flipped. The process is time-reversion symmetric.
}
\end{center}
\end{figure}

The induced metric on the worldsheet has components
\bea
\nonumber
 & g_{+-} = -\le(1-{A^2 \ov 2} \tanh \epsilon\sigma^- \tanh \epsilon \sigma^+ \ri)^2 & \\
\nonumber
& g_{++} = g_{--} = 0 &
\eea
The Ricci scalar of the induced metric is
\be
  \nonumber
  R = - {2 A^2 \epsilon^2 \over (1- {A^2 \ov 2} \tanh \epsilon \sigma^- \tanh \epsilon \sigma^+ )^4 (\cosh \epsilon \sigma^-)^2 (\cosh \epsilon \sigma^+)^2  }
\ee
Integrating this with respect to $\sigma^-$ and $\sigma^+$ over the entire worldsheet, we get
\be
  \nonumber
  \int d^2\sigma \sqrt{-g} \, R = -{16} \tanh^{-1} {A^2 \over 2}
\ee

Note that $\epsilon$ has dropped out and thus the result is finite in the $\epsilon \to \infty$ limit that corresponds to sharp kinks. In this limit, $R=0$ away from the collision vertex at $\sigma^+ = \sigma^- = 0$. Thus, it is enough to integrate in an infinitesimally small neighborhood of the origin. The result is then a local feature which generalizes to AdS$_3$.

We would like to eliminate $A$ from the expression and replace it with a  more natural quantity. The angle between the two static string pieces can be computed
\be
  \nonumber
  \tan {\phi \ov 2} = {2 \sqrt{2} A \over A^2 - 2}
\ee
Going back to AdS$_3$, we can set up a similar collision using the three patches $N_1$, $N_2$, $N_3$
\bea
  \nonumber
  \vec N_1 &=& \le(\quad 0, \qquad 0, \quad \cos{\phi \over 2}, \quad \sin{\phi \over 2} \ri) \\
  \nonumber
  \vec N_2 &=& \le(-\tan{\phi \over 2}, \, 0, \, \le(\cos{\phi \over 2}\ri)^{-1}, \, 0 \,\,\, \ri) \\
  \nonumber
  \vec N_3 &=& \le(\quad 0, \qquad 0, \quad \cos{\phi \over 2}, \, -\sin{\phi \over 2} \ri)
\eea
The angle $\phi$ can be computed using the scalar product between normal vectors
\be
  \nonumber
  \cos \phi = \vec N_1 \cdot \vec N_3
\ee
and from this, $A$ can be determined. The final result
\be
  \nonumber
     \int_\textrm{vertex} \sqrt{-g} \, R =   8 \log \cos {\phi \ov 2}
\ee
This is the integrated Ricci scalar around a kink collision point where the string piece corresponding to the middle patch $N_2$ vanishes.
Note that the formula does not depend on $N_2$. Generic normal vectors form a three-dimensional space. However, the middle patch $N_2$ is constrained by  $\vec N_1 \cdot \vec N_2 =\vec N_2 \cdot \vec N_3 = 1$ and thus the allowed values  form a one-dimensional subspace. Motion on this subspace corresponds to global AdS$_3$ time translations. This is a symmetry of the system that preserves the curvature.

The four angle variables around the vertex in the kink lattice are:
\be
  \nonumber
  \alpha_{1} = -\alpha_{3} = {\phi \ov 4}, \quad \alpha_{2}  ={\pi\ov 2} - {\phi \ov 4},
  \quad \alpha_{4}  ={\pi\ov 2} + {\phi \ov 4}
\ee
If $\phi=0$, then the angles do not change at the vertex (i.e. $\alpha_1 = \alpha_3$ and $\alpha_2 = \alpha_4$) and it's clear that they describe the collision of two zero kinks.


\end{document}